\documentclass[sigconf]{acmart}

\usepackage{graphicx}
\usepackage{algorithm}
\usepackage{algorithmic}
\usepackage{subfigure}
\usepackage{array}
\usepackage{setspace}
\usepackage{float}
\usepackage{url}
\usepackage{balance}
\usepackage{amsfonts}
\usepackage{amsmath}
\usepackage{rotating}
\usepackage{multirow}
\usepackage{color}
\usepackage{bm}
\usepackage{enumitem}
\usepackage{pifont}
\usepackage[T1]{fontenc}
\usepackage{colortbl}
\usepackage{tabularx}
\usepackage{tablefootnote}
\definecolor{darkg}{rgb}{0.0, 0.5, 0.0}

\copyrightyear{2021}
\setcopyright{iw3c2w3}
\acmConference[]{}
\acmPrice{}
\acmDOI{}
\acmISBN{}

\begin{document}
\title{Is preprint the future of science? \\ 
 A thirty year journey of online preprint services}

\author{Boya Xie}
\affiliation{
\institution{Microsoft Research, Redmond}
\institution{boxie@microsoft.com}
}

\author{Zhihong Shen}
\affiliation{
\institution{Microsoft Research, Redmond}
\institution{zhihosh@microsoft.com}
}

\author{Kuansan Wang}
\affiliation{
\institution{Microsoft Research, Redmond}
\institution{kuansanw@microsoft.com}
}

\begin{abstract}
  Preprint is a version of a scientific paper that is publicly distributed preceding formal peer review. Since the launch of arXiv in 1991, preprints have been increasingly distributed over the Internet as opposed to paper copies. It allows open online access to disseminate the original research within a few days, often at a very low operating cost. This work overviews how preprint has been evolving and impacting the research community over the past thirty years alongside the growth of the Web. 
  In this work, we first report that the number of preprints has exponentially increased 63 times in 30 years, although it only accounts for 4\% of research articles. Second, we quantify the benefits that preprints bring to authors: preprints reach an audience 14 months earlier on average and associate with five times more citations compared with a non-preprint counterpart. Last, to address the quality concern of preprints, we discover that 41\% of preprints are ultimately published at a peer-reviewed destination, and the published venues are as influential as papers without a preprint version. Additionally, we discuss the unprecedented role of preprints in communicating the latest research data during recent public health emergencies. In conclusion, we provide quantitative evidence to unveil the positive impact of preprints on individual researchers and the community. Preprints make scholarly communication more efficient by disseminating scientific discoveries more rapidly and widely with the aid of Web technologies. The measurements we present in this study can help researchers and policymakers make informed decisions about how to effectively use and responsibly embrace a preprint culture. 
\end{abstract}

\keywords{preprint analysis, citation impact, publication rate, time to publish, arXiv, SSRN, bioRxiv}

\begin{spacing}{1}

\maketitle

\begin{figure}[!t]
  \centering
  \includegraphics[width=10cm]{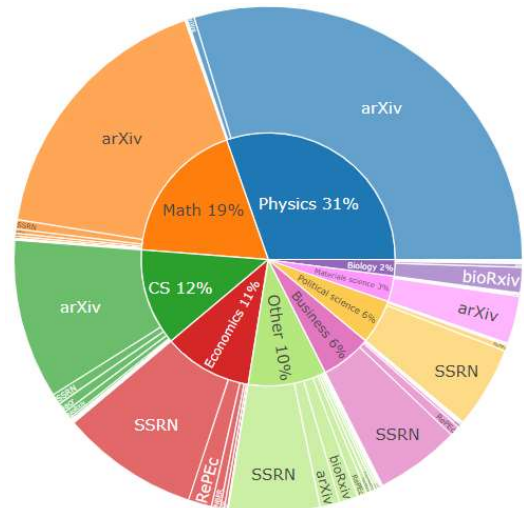}
  \vspace{-0.5cm}
  \caption{Overall preprints distribution}
    \label{fig:fosall}
\end{figure}

\section{Introduction}

Preprint is a form of a scholarly article which is not peer-reviewed yet but made available either as paper format or electronic copy. If the curiosity and excitement to push forward the frontier of collective human knowledge makes scientists desire to share their discoveries as soon as possible to as wide audiences as possible, then online preprint could be the ultimate manifestation of such spirit of science: accessible for anyone with the Internet access shortly after the paper is uploaded. In the past thirty years, the Web deeply impacts all aspects of our lives and the society, including the academic publishing. In this work, we study the recent evolution of preprint from 1991 to 2020 to understand whether the permeated Web technology could help re-establish a broader adoption of the preprint tradition for a more efficient and effective dissemination of science.

Preprints were distributed through physical mails and had been struggled with high postage cost for decades until Paul Ginsparg at the Los Alamos National Laboratory launched an automated email server distributing high energy physics preprints in 1991 \cite{Garisto19,Rosen1970,Ginsparg2011}. The email server later evolved to a web service known as arXiv, hosting preprints in eight different fields such as physics, mathematics, computer science and quantitative biology. Today arXiv hosts more than 1.7 million articles - one of the largest preprint servers in size. In 1994, the most popular repository for economics, law and the social sciences, Social Science Research Network (SSRN) was founded by two economists \cite{Noorden2016}. Subsequently in 1997, another economics preprint server, Research Papers in Economics (RePEc) was created \cite{RePEc}. While just a small number of preprints launched from 2000 to 2010, more than 40 preprint servers have been created in the past decade \cite{Johnson2019}. Some are general-purpose, such as PeerJ PrePrints, Preprints.org, viXra; some are for specific disciplines, such as bioRxiv, medRxiv, LawArXiv, SportRxiv; and some are for specific regions or languages, such as AfricArXiv, Arabixiv, IndiaRxiv, and Frenxiv. Figure \ref{fig:fosall} shows the preprint distribution in different disciplines and servers hosting the articles.

In this work, we employ a large-scale multi-disciplinary preprint dataset sourced from the Microsoft Academic Graph (MAG) \cite{Sinha-15} to explore the trends and impacts of preprint quantitatively. This dataset includes 2.8 million preprints and 69.8 million peer-reviewed articles from 1991 to 2020 and it is the largest preprint dataset used for analysis to the best of our knowledge. 

\begin{itemize}[leftmargin=*]

\item \textbf{Trends and adoption of preprint.} To understand how different research community and disciplines adopt the preprint culture,  %\textbf{Section ~\ref{sec:trend}} 
we analyzes the development of preprints over the past thirty years. The annual number of preprints has increased from a few thousands in 1991 to 227k in 2019, although it still accounts for a small portion of all scientific papers. In the first nine months of 2020, 192k preprints are produced, which represent 6.4\% of all papers published in the same period. When zooming in to individual domains: \textit{physics}, \textit{mathematics}, and \textit{economics} adopt the preprint culture the best; increasing popularity is observed in \textit{computer science} and \textit{biology} in the last 5 years; and preprint is rare in other domains. 

\item \textbf{Impacts on individual researchers}. The data shows when a work is documented in preprint, it reaches an audience 14 months earlier than a peer-reviewed counterpart. We also unveil that the early release is associated with two times more citations for a paper. 

\item \textbf{Impacts on research communities}. One of the biggest concern of adopting preprint is the quality without a peer endorsement system in place. In this analysis, we find preprints have sufficient quality control and the paper quality is similar to peer-reviewed publications. We discover that 41\% of preprints ultimately have passed peer review and are published in journals or conferences, and the destination venues are more influential than the venues of peer-reviewed papers. Preprint shows significant advantage in timely sharing of the latest findings during COVID-19 pandemic and it plays an unprecedented role in the dissemination of biomedical and life science knowledge at such public health emergencies. 
\end{itemize}

Overall, our findings demonstrate that with the penetration of Web technologies, online preprint servers enable early release and attract more citations for individual researchers. Preprints also benefit the research community by hosting adequate-quality papers with prompt open access. These benefits might be overlooked before due to the lack of quantitative measurements. The endeavour of scientific communication aims to spread knowledge as rapidly, as widely, and as cost-effectively as possible. This work provides evidence that we could achieve this goal better at this Web era with a responsible adoption of the preprint culture and, ultimately, to collectively advance the frontier of human knowledge and make it available to public usage more effectively.

\section{Data and Method}
In order to have a holistic view of online preprint and its evolution in different disciplines, we design our analysis to include as many major preprint servers as possible, and cover fields from natural science to social science. MAG as a Web-scale scholarly dataset serves the purpose well. This section describes how the preprint dataset is generated from MAG, and what are the assumptions and known limitations.

\subsection{Microsoft academic graph} 
MAG is a heterogeneous graph containing scientific publications, citation relationships, as well as other academic entities such as authors, institutions, journals, conferences, and fields of study. The graph is updated weekly to include newly released scholarly articles on the Web. As of October 2020, there are over 240 million scientific documents from 1800 to 2020 in MAG with eight different types: journal, conference, book, patent, book chapter, repository, thesis, and other.\footnote{All data presented in this study are based on MAG 2020-10-01 version.} The preprint servers covered in this work are listed in Table \ref{tab:preprintlist}. 

Totally 2.8 million preprints are found in MAG based on document types and URL domains, among which 1.7 million are contributed by arXiv, followed by SSRN. SSRN claims to host 0.95 million abstracts and 0.82 million full text papers \cite{SSRNStats}. However MAG only contains 0.76 million preprints sourced from SSRN. We notice there is a significant drop in the number of SSRN papers posted online after 2016 in MAG, coincident with Elsevier's acquisition of SSRN. Since we don't find any other sources providing downloadable SSRN data, MAG SSRN data post 2016 is treated as incomplete, related evaluations are affected accordingly. Besides SSRN, the collections for Keldysh Institute preprint, PsyArXiv, and SocArXiv are also incomplete in MAG. There are other smaller preprint servers not being covered in this study, but we believe the major players in the preprint services are included and the data is sufficient to unveil the key characteristics of the preprint landscape.

Online preprints started in 1991 with the launch of arXiv server. There are less than 1\% preprints prior to 1991 in MAG. Therefore this study only focuses on preprints starting from 1991. In order to compare preprints with peer-reviewed papers, journal and conference papers from 1991 to 2020 are also included. We exclude other types of documents (e.g. technical reports) with the assumption that a vast majority of journal and conference papers are peer-reviewed while other types of documents may not go through a rigorous peer endorsement process. All analysis performed in this study are based on past 30 years data unless otherwise specified. 

\begin{table}[!t]
  \begin{tabular}{c|c|c|c}
    \toprule
    Server&\multicolumn{1}{|p{1.7cm}}{\centering Self-reported size}&\multicolumn{1}{|p{1cm}|}{\centering Size in MAG}&Year range\\
    \midrule
   arXiv & 1,777,025 \tablefootnote{https://arxiv.org/ accessed Oct. 14, 2020 }  & 1,753,132 &  1992-present\\
SSRN & 950,733\tablefootnote{https://papers.ssrn.com/sol3/DisplayAbstractSearch.cfm accessed Oct. 14, 2020} & 759,075 & 1994-present\\
RePEc & N.A. & 100,284 & 1993-present\\
bioRxiv &  98,301\tablefootnote{ \url{http://api.biorxiv.org/reports/content_summary}}  & 100,109 & 2013-present \\
viXra & 35,893 \tablefootnote{https://vixra.org/ accessed Oct. 14, 2020} & 31,497 & 2007-present\\
NBER Working Paper & 24,000+\tablefootnote{\url{https://www.nber.org/pubs.html\#:~:text=The\%20NBER\%20working\%20paper\%20series,includes\%20more\%20than\%2024\%2C000\%20papers.}} & 31,128 & 1973-present\\
IACR ePrint & N.A.  & 23,291 & 1996-present\\
Preprints.org & 16,951\tablefootnote{https://www.preprints.org/ accessed Oct. 14, 2020} & 17,337 & 2016-present\\
medRxiv & N.A. & 11,568 & 2019-present\\
IMF Working Paper & 7,008\tablefootnote{\url{https://www.imf.org/en/publications/search?when=After&series=IMF+Working+Papers} accessed Oct. 14, 2020} & 8,016 & 1986-present\\
Peerj preprints & 5,068\tablefootnote{https://peerj.com/preprints/} & 4,066 & 2012-2019\\
ACS PETR & N.A.  & 3,028 & 1920-2012\\
Nature Precedings & N.A. & 2,986 & 2007-2012\\
Polymer Preprints & N.A. & 1,214 & 1966-2014\\
Keldysh Preprint & 2,849\tablefootnote{\url{http://www.mathnet.ru/php/journal.phtml?jrnid=ipmp&option_lang=eng} accessed Oct. 14, 2020} & 797 & 1967-present\\
EconStor Preprints  & 713\tablefootnote{\url{https://www.econstor.eu/handle/10419/30002} accessed Oct. 14, 2020} & 721 & 1986-present\\
SocArXiv  & 6,439\tablefootnote{https://osf.io/preprints/socarxiv accessed Oct. 14, 2020} & 564 & 2016-present\\
chemRxiv &  N.A.  & 307 & 2017-present\\
PsyArXiv & 11,959\tablefootnote{https://psyarxiv.com/ accessed Oct. 14, 2020} & 28 & 2016-present \\
  \bottomrule
\end{tabular}
  
\vspace{2mm}
\caption{Online preprint servers covered in this study}
  \label{tab:preprintlist}
\vspace{-6mm}
\end{table}

\subsection{Experiment design}
\begin{figure}[h]
  \centering
  \includegraphics[width=.8\linewidth]{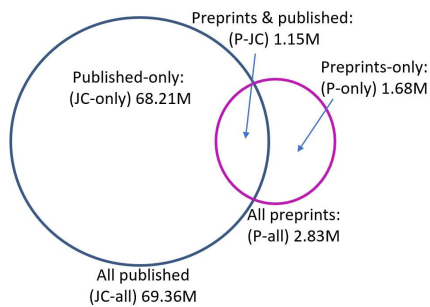}
  \vspace{-0.5cm}
  \caption{Paper groups and sizes used in this analysis} \label{fig:preprint-size}
\end{figure}
To facilitate evaluations, we split preprints, journal and conference papers from 1991 to 2020 into three mutually exclusive groups (P-only, JC-only, P-JC) based on whether they are submitted to preprints or are published at a peer-reviewed journal or conference. In total, five groups are defined below and we summarize their relationships and sizes in Figure~\ref{fig:preprint-size} and their usages in subsequent analysis in Table \ref{tab:experiment}.
\begin{itemize}[leftmargin=*]
  \item P-only: papers only submitted to preprints;
  \item JC-only: papers only published;
  \item P-JC: papers submitted to preprints and published;
  \item P-all (P-only + P-JC): all preprints;
  \item JC-all (JC-only + P-JC): all papers published. 
\end{itemize}

\subsection{Assumptions and limitations}
\textbf{Preprint and post-print.} Authors normally upload papers to preprint servers either without submitting to a journal, or while waiting for the peer review result or for the official publishing. However, there are also cases that papers are uploaded to repositories after they are published in journals or conferences, which are called post-prints. Some servers restrict the submission to pre-publication only, while others host both. ArXiv and SSRN allow both preprint and post-print. Since there isn't a clear signal to differentiate post-prints from preprints, and we estimate\footnote{Post-print is assumed if the paper's publication date is prior to the preprint online date.} there are about 4\% post-print in arXiv and 2\% in SSRN, we treat all articles in arXiv and SSRN as preprints in this study. 

\begin{table}[!h]
%  \begin{tabular}{c|c|c|c}
\begin{tabularx}{\linewidth}{c|c|X|X}
    \toprule
    Section&Figure&Usage&Measurement\\
    \midrule
   1 \& 3.2 & 1 & Preprint distribution & P-all\\
3.1 & 3 & Trend & P-all and P-all/(P-all + JC-all)\\
3.2 & 4 & Trend by domain & P-all/(P-all + JC-all)\\
4.1 & 5 & Days to publish by server & P-JC(publishDate - onlineDate)\\
4.1 & 6 & Days to publish by domain & P-JC(publishDate - onlineDate)\\
4.2 & 7 & Citation impact & P-JC.citation and JC-only.citation\\
4.2 & 8 & Citation impact by domain & P-JC.citation - JC-only.citation\\
5.2 & 9 & Publish rate by domain & P-JC/P-all\\
5.2 & 10 & Publish rate by server & P-JC/P-all\\
5.3 & 11 & IF distribution & JC-all and P-JC\\
5.3 & 12 & IF distribution by domain & JC-all and P-JC\\
  \bottomrule
\end{tabularx}
  
\vspace{2mm}
\caption{Online preprint servers covered in this study}
  \label{tab:experiment}
\vspace{-6mm}
\end{table}

\textbf{Preprint and working paper.}
Multiple economics paper servers, such as RePEc, NBER, and IMF, refer to unpublished articles as working papers. A working paper shares many things in common with preprints: both are non-peer-reviewed, both are made available online preceding publication to reach audiences earlier. Sometimes working paper could be preliminary results and less mature than preprint, and other times working paper and preprint are used interchangeably. Here we treat working papers from the above mentioned three servers as preprints.

\textbf{Paper discipline categorization.} To perform the analysis on preprints in individual disciplines, we leverage the domain categorization of each paper in MAG~\cite{shen2018a}. MAG has a high-quality scientific concept (or field of study) taxonomy constructed semi-automatically. As of October 2020, there are more than 740K concepts organized into a six level hierarchy with 19 top level domains, such as: \textit{biology}, \textit{computer science}, \textit{economics}, etc. Each paper is assigned with at least one top level domain. Interdisciplinary papers are labeled with multiple top level domains, e.g. a biochemical paper is labeled with both \textit{biology} and \textit{chemistry}. For all domain-related analysis, interdisciplinary papers are counted once towards each of its belonging domains. \textit{Physics}, \textit{mathematics}, \textit{computer science}, \textit{economics}, and \textit{biology} are selected for disciplinary analysis because they either have the most number of preprints or are fast evolving.

\section{Preprint Trend}\label{sec:trend}
In this section, we provide an overview of the preprint general trends and development in different disciplines. The annual preprint volume and ratio are evaluated.

\subsection{Overall trend}
\begin{figure}[h]
  \centering
  \includegraphics[width=\linewidth]{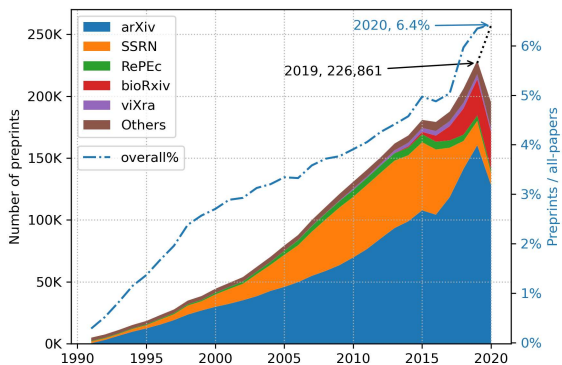}
   \vspace{-1cm}
  \caption{Annual number of preprints (P-all) and reprints(P-all)/all-papers(P-all+JC-all) rate growth}
\label{fig:trend}
\end{figure}

The number of preprints uploaded online per year has been increasing steadily in the past decades, from 3k in 1991 to 227k in 2019. There are at least 192k new preprints made available online in the first nine months of 2020. Previously, the volume of overall scientific literature are found to be growing exponentially and double every 15 years on average \cite{Santo2018}. The preprint follows a similar growth trajectory but doubles in less than 10 years. Despite the increasing number, the ratio of preprints to all scientific articles is still very low, only 6.4\% in 2020, as shown in Figure \ref{fig:trend}. ArXiv is a dominant player in preprint services, where its change is illustrated in the bottom area of the figure. The volume growth in arXiv defines the overall trend of all preprints. As described in the previous section, SSRN volume is expected to be higher post 2015. BioRxiv becomes an important contributor starting from 2015 as shown in the red area. It has generated the third largest volume in the past five years.

\subsection{Trend by domain}
\begin{figure}
  \centering
  \includegraphics[width=\linewidth]{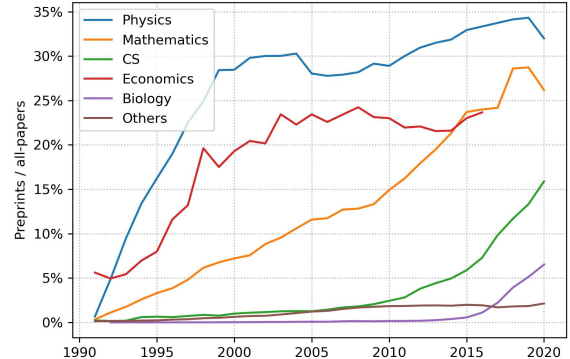}
    \vspace{-1cm}
  \caption{Yearly preprints(P-all)/all-papers(P+JC) trend by domain}
      \vspace{-0.5cm}
\label{fig:fostrend}
\end{figure}

The development of preprint is different in each domain: \textit{physics} and \textit{mathematics} are the pioneers as people in these communities are more intimate with Web technologies since early days; some are new active players within the past 10 years, such as \textit{computer science} and \textit{biology}. Figure \ref{fig:fosall} shows the preprint distribution in various disciplines and servers hosting the articles. Among all preprints, half are physics and mathematics papers, where 96\% of physics and mathematics preprints are uploaded to arXiv. Considering the history of online preprints, it is not surprising to find that 31\% of all preprints are physics papers. Publishing papers as a preprint is the norm in physics. Figure \ref{fig:fostrend} exhibits that three out of ten physics papers were preprints in the past twenty years. On the other hand, \textit{mathematics} is where we see the second largest volume, and there has been increasing level of interest of using preprints from mathematicians. Although \textit{computer science} has the third largest number of preprints, majority (95\%) of computer science papers are non-preprints due to the high throughput of computer science literature. Besides physics and mathematics which have a high ratio of preprints to all-papers at 27.7\% and 15.7\% respectively, \textit{economics} has a relatively high average ratio at 20.8\% from 1991 to 2015. SSRN hosts the most economics preprints, as well as other social sciences such as business and political science, shown in Figure \ref{fig:fosall}. While many other disciplines have less than 2\% preprints, \textit{biology} has been promoting preprint since 2013, and a good momentum is observed in the past five years. Biology has 6.5\% preprints in 2020 and majority of them are in bioRxiv.

\section{Impact on individual researcher}\label{sec:individual}
Multiple articles discussed the concerns that authors might have over submitting preprints: scooping, being precluded from journal publication, and low quality implication, etc.\cite{Bourne2017, Kaiser2017, Clemens, Sarabipour2019}. On the other hand, the benefits of preprints are frequently mentioned as well. Since it is not straight-forward to design quantitative measurements from the publishing data of above mentioned pitfalls. Instead, we exam some claimed benefits of preprints for individual researchers.

\subsection{Preprints provide early visibility}
Presenting scientific findings through paper puts a timestamp on the work. The lengthy peer review and publishing process could take months or even years. On the contrary, uploading a preprint paper to get a permanent open accessed timestamp only takes a few days. In addition to this unbeatable advantage of time-stamping a paper, preprints enable the work to receive credits (in the form of citations) faster, which is crucial for young researchers whose career progression heavily  depends on a timely recognition of their work \cite{Sarabipour2019}.
\begin{figure}[h]
  \centering
  \includegraphics[width=\linewidth]{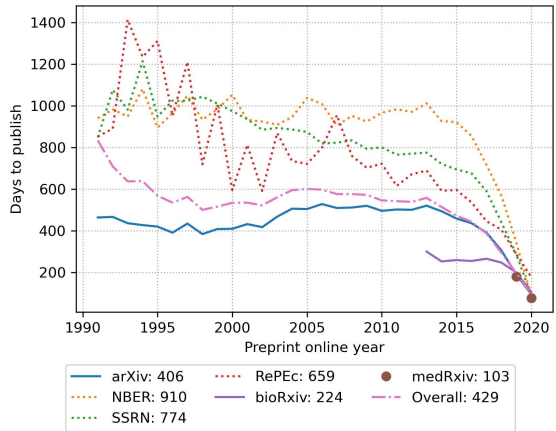}
  
\vspace{-2mm}
  \caption{Days to publish  by preprint server}
\label{fig:days2pub}
\end{figure}

\begin{figure}[h]
  \centering
  \includegraphics[width=\linewidth]{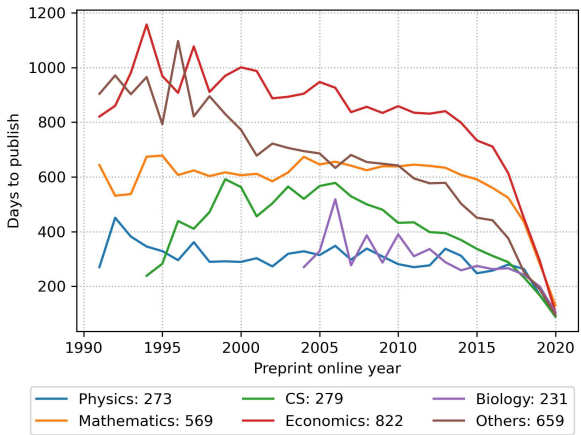}
  \vspace{-1cm}
  \caption{Days to publish by domain}
      \vspace{-0.5cm}
\label{fig:days2pubfos}
\end{figure}
Here, we analyze the duration between preprints' initial online date and the journal/conference official publication date to evaluate how much the preprint expedites an article to reach an audience to gain the potential earlier recognition. The analysis is conducted from two perspectives: days difference in each preprint server, and with different disciplines. Figure \ref{fig:days2pub} exhibits the measurement for six major preprint servers and the average of all preprints. The average days to publish of different preprint servers are listed in the legend of Figure~\ref{fig:days2pub}. ArXiv has similar days to publish to the overall average of all servers. Both are around 400 days. 
It takes much longer to publish for papers from NBER, SSRN, and RePEc (shown with dotted lines in Figure ~\ref{fig:days2pub}), which is 
0.5 to 1.1 times longer compared to the average. These three servers are where most economics preprints submit to. This result demonstrates a huge benefit that preprints bring to economics researchers for earlier visibility. We assume that a first version preprint is similar to a ready-for-peer-review paper, and both take similar amount of time to be published in peer-reviewed destinations. Figure \ref{fig:days2pubfos} implies economics papers take the longest time to publish, at 822 days. This finding is consistent with the observation of a prolonged peer review process in economics~\cite{Moore1965,Ozler2011}. By submitting to preprint servers, an economic researcher can get recognition 2.25 years earlier. Biology has the shortest lag, about 7 months. Figure \ref{fig:days2pub} shows preprints in medRxiv only take 103 days to publish. However, medRxiv just launched in June 2019. Many papers in medRxiv might still undergo the review and publishing process right now thus being excluded from the plots in Figure \ref{fig:days2pub}. The measurement for medRxiv will be more accurate when there is more data available over time to overcome the current bias from the early published "survivors".

To conclude, preprints help papers become accessible 7 months to 2.25 years earlier than peer-reviewed counterparts, and economics researchers would benefit from it the most.

\subsection{Preprints correlate with more citations}
The early visibility provided by preprints generates opportunities for scientific work to make more impacts. One of the most direct and recognized measurement of research impact is paper citations. The citation is found to follow power-law distribution \cite{Golo2012, stege2012, Waltman2012} and the dynamics is universal for all disciplines \cite{Santo2018}. The preprint is no exception, Appendix \ref{appendix:citdis} shows the preprint citation follows a power-law distribution. To illustrate whether preprints is associated with more citations, we compare the papers' citation of published papers with (P-JC) and without (JC-only) a preprint version.

\begin{figure}[h]
  \centering
  \includegraphics[width=\linewidth]{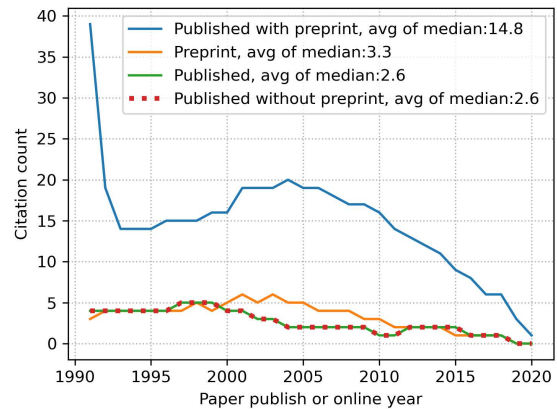}
  \vspace{-1cm}
  \caption{Overall citation impact}
   % \vspace{-0.5cm}
\label{fig:citeall}
\end{figure}
 The yearly median citations of the four groups of papers: P-JC, JC-only, P-all, and JC-all, are illustrated in Figure \ref{fig:citeall}. It manifests that papers with preprints have more citations in any year regardless whether they have a published version. On average of all years, a journal or conference paper with a preprint version (P-JC) has median citation of 14.8, while a non-preprint counterpart (JC-only) receives 2.6 citations, which results in 12.2 citation difference (five times more). All preprints (P-all), regardless whether they have been published in journal/conference, have 0.7 more citations than papers without a preprint version (JC-only). 

\begin{figure}[h]
  \centering
  \includegraphics[width=\linewidth]{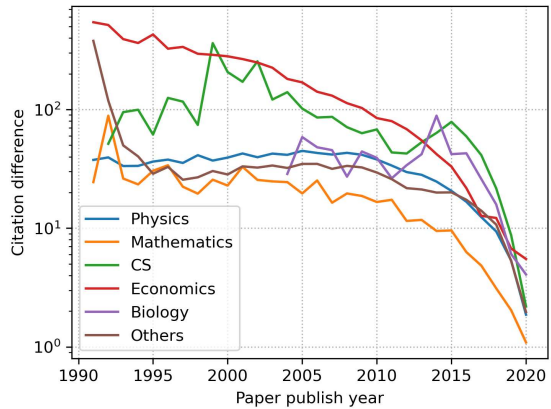}
  
\vspace{-2mm}
  \caption{Citation impact by domain (citation of published paper with preprint (P-JC) subtract citation of published paper without preprint (JC-only))}
\label{fig:citefos}
\end{figure}

Figure \ref{fig:citefos} demonstrates the citation difference between papers with and without preprints in different domains. The citation difference is defined as the yearly average citation per published paper with preprints (P-JC) subtracting the yearly average citation per paper without preprints (JC-only). The citation difference is positive and it indicates that  published papers with a preprint version (P-JC group) have bigger impacts as the citation of P-JC group is more than the citation of JC-only group. Here we reveal that preprint has positive citation impact in all domains, in which economics has the most citation difference (6.9-fold more) and mathematics has the least (0.95-fold more). The number of citations and citation differences for each discipline are listed in Table \ref{tab:citimpact}.

\begin{table}[h]
    \centering
    \begin{tabular}{c|c|c|c}
        \toprule
    Domain&\multicolumn{1}{|p{2cm}}{\centering Paper with preprint citations (P-JC)}
    &\multicolumn{1}{|p{2cm}}{\centering
    Paper without preprint citations (JC-only)}
    &\multicolumn{1}{|p{1.5cm}}{\centering Citation ratio (P-JC/JC-only)} \\
    \midrule
    Economics & 220 & 28 & 7.86 \\
CS & 116 & 21 & 5.52 \\
Physics & 50 & 18 & 2.78 \\
Others & 72 & 31 & 2.32 \\
Biology & 68 & 31 & 2.19 \\
Mathematics & 43 & 22 & 1.95 \\
  \bottomrule
  Overall&74&24&3.08\\
  \hline
    \end{tabular}
  
\vspace{2mm}
    \caption{Citation impact}
    \label{tab:citimpact}
    
\vspace{-6mm}
\end{table}

By evaluating 69.4 million papers published from 1991 to 2020 over all disciplines, we uncover that preprints have significant positive contribution to paper citations. This result is consistent with previous studies\cite{Serghio2018, Beccot2009} at a dataset which is hundreds times larger with more comprehensive coverage on all disciplines and a longer time span.

In this section, we quantify the positive impacts that preprints bring to individual researchers in terms of days to publish and the citation difference. Preprints make papers available to public 14 months earlier compared to peer-reviewed papers. It is most beneficial to economic researchers who usually undergo a prolonged publishing process up to 3 years. Preprints correlate with two times more citations on average, and always have positive effect on citations of papers in all domains that we evaluate.

\section{Impact on the research community}\label{sec:community}
Is preprint contributing to the research community similar to peer-reviewed articles or is it diluting the novel findings with inaccurate or even incorrect conclusions? Do preprints serve better in certain scenarios where peer-reviewed journals or conferences have limitations? In this section, we address the quality concerns of preprints by first reviewing and summarizing the established ways of quality control on preprint servers. We then assess the population of preprints ultimately being published in a peer-reviewed venue and evaluate the impact factors of these venues to further provide the quantitative evidence of the preprints' quality. At last, we review and discuss the important role of preprints in recent public health emergencies and how preprints help to speed up the dissemination of biomedical knowledge in a global pandemic. 

\subsection{No peer review is not equivalent to no quality control}
Preprints would not be valuable to the research community if the content were substandard. The quality of preprints is one of the major concerns of preprint consumers as there is no formal peer review process in place. However there are other processes implemented at different servers to help the quality control.

\begin{itemize}[leftmargin=*]
\item \textbf{Author control:} arXiv requires author registration and endorsement for new users \cite{ArxivEndor}; IMF working paper \cite{IMF} and Keldysh Institute Preprints \cite{Keldysh} only include works from its own employees or members.
\item\textbf{Screening process:} many preprint servers perform checks on completeness, relevance, plagiarism, appropriate language and contents, as well as the ethical and legal compliance \cite{ArxivSub, biorxivSub, SSRNSub, Kirkham2020}.
\item \textbf{Extended content review:} bioRxiv and medRxiv reject materials that might pose a health or biosecurity risk \cite{biorxivSub, medrxivSub}. Additionally preprint platforms are taking extra cautions when the paper might have an immediate impact on public health. ArXiv and ChemRxiv have enhanced their screening on COVID-19 related articles and bioRxiv has stopped accepting articles "making predictions about treatments for COVID-19 solely on the basis of computational work"~\cite{Kwon2020}.
\end{itemize}

The peer review process functions as an error and fraud capturing tool, therefore peer-reviewed papers are perceived as quality guaranteed today. Oftentimes, the value of a paper is linked to the prestige of the publication venue, especially when it comes to grant application or career progression. Nevertheless, peer review has its own limitations. Besides issues such as sustainability, the peer review fraud, unfair reviews, and incompetent reviewers~\cite{Alberts2008, Stahel2014, Fountain2014}, several studies have shown that peer reviews do not always detect errors, and some peer-review-rejected research works are not necessarily low quality, some are even Nobel Prize winning discoveries~\cite{Smith2006, Allison2016, Balietti2016, Xiao2020}. 

\subsection{Adequate amount of preprints do pass peer review}
\begin{figure}[h]
  \centering
  \includegraphics[width=\linewidth]{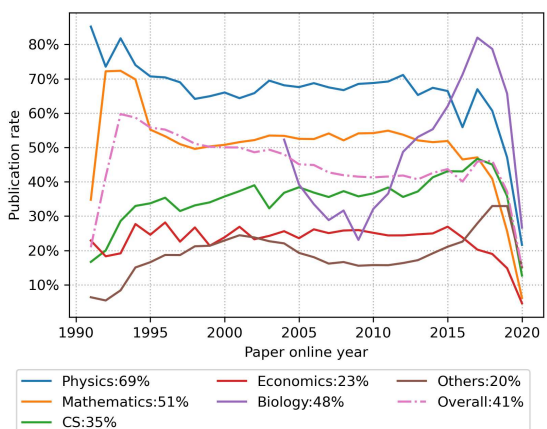}
  \vspace{-1cm}
  \caption{Preprint publication rate (P-JC/P-all) by domain}
  \vspace{-0.5cm}
  \label{fig:pubratefos}
\end{figure}

\begin{figure}[h]
  \centering
  \includegraphics[width=\linewidth]{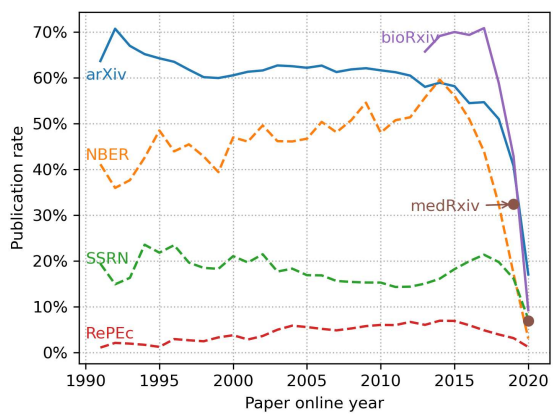}
  \vspace{-1cm}
  \caption{Preprint publication rate (P-JC/P-all) by server}
    \vspace{-0.5cm}
  \label{fig:pubratevenue}
\end{figure}

Put the criticisms of peer review aside, publishing at a well branded journal or conference is still one of the most common ways to certify the quality of an article. Evaluating 2.8 million preprints in MAG, we find that 41\% of them eventually are published at peer-reviewed venues, shown in Figure \ref{fig:pubratefos}. \textit{Physics} and \textit{mathematics} are the top two domains having most preprints published in journals or conferences, 69\% and 51\% respectively. Comparably stable publication rates are observed in all domains except \textit{biology}, which jumps from 23\% in 2009 to over 80\% in 2017. Notice the new player bioRxiv has the highest publication rate among all preprint servers shown in Figure \ref{fig:pubratevenue}. The publication rate of bioRxiv in our result is consistent with the value reported by Abdill and Blekhman~\cite{Abdill2019}. BioRxiv is the major contributor to the elevated preprint publication rate in biology in recent years. MedRxiv is a new medical preprint server launched by the same founder as bioRxiv, Cold Spring Harbor Laboratory, in June 2019. Within one year, medRxiv already has 32\% of papers published. This number is likely to grow higher over time as some papers might be waiting to be published. The three dashed lines in Figure \ref{fig:pubratevenue} represent the publication rate for three most used economics preprint servers. The NBER has the best rate among the three, close to arXiv. SSRN comes second, but has less than half of the NBER publication rate. RePEc have further more portions of preprints stay unpublished, with less than 10\% published.

\subsection{Preprint is published in venues with higher impact factor}
\begin{figure}[h]
  \centering
  \includegraphics[width=\linewidth]{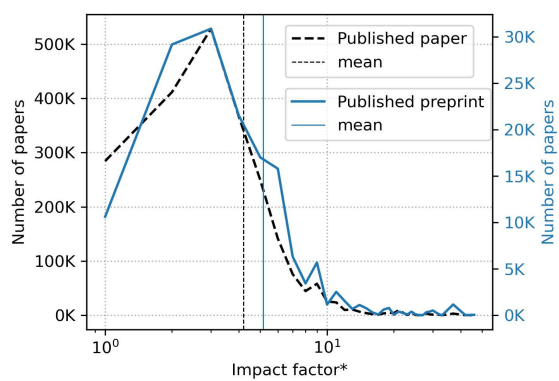}
  \vspace{-1cm}
  \caption{17'-18' published paper (JC-all) and published preprint (P-JC) distribution over venue impact factors ( * calculated with MAG data)}
  \vspace{-0.5cm}
  \label{fig:impact}
\end{figure}

\begin{figure}[h]
  \centering
  \includegraphics[width=\linewidth]{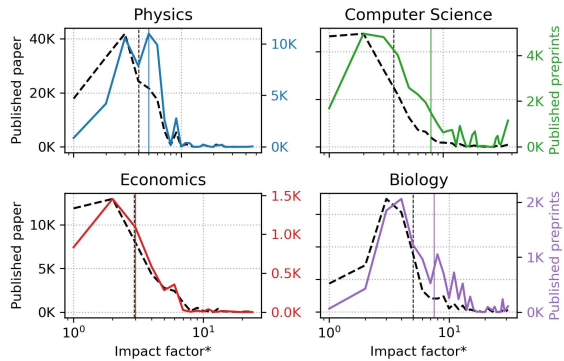}
  \vspace{-1cm}
  \caption{17'-18' published paper (JC-all) and published preprint (P-JC) distribution over venue impact factors by domain ( * calculated with MAG data)}
  \vspace{-0.5cm}
  \label{fig:impactfos}
\end{figure}

With the cohort of published preprints (J-PC group), we further analyze their destination journals and conferences. To check whether preprints tend to be accepted by less influential venues, we use the 2019 two-year impact factor as the impact indicator and plot the number of published preprints (P-JC) and the number of all published papers (JC-all) with respect to their destination impact factors in Figure \ref{fig:impact} and \ref{fig:impactfos}. The 2019 impact factor is calculated as the average citation received in 2019 for papers published in 2017 and 2018~\footnote{https://en.wikipedia.org/wiki/Impact\_factor} using MAG data:
\begin{displaymath}
    IF_{2019}=\frac{Citations_{2019}}{Publication_{2018}+Publication_{2017}}
\end{displaymath}
Since the calculation of the impact factor is based on publications in 2017 and 2018, only papers published in these two years account for this distribution analysis. There are 2,278,117 papers published in 2017 and 2018, and among them 152,799 papers have a preprint version. Figure \ref{fig:impact} illustrates the paper distributions over venue impact factors, where solid blue line denotes published preprints, i.e. P-JC papers published in 2017 and 2018, and dashed black line represents all papers, i.e. JC-all papers published in 2017 and 2018. The distributions of the two groups are closely aligned with each other and both are right skewed with similar means. The published preprints are accepted by venues with higher impacts (impact factor 5.2 and 4.2 for P-JC and JC-all groups respectively). When we take a closer look at different domains, in \textit{physics}, \textit{computer science}, and \textit{biology}, published preprints shift more towards higher impact factors in Figure \ref{fig:impactfos}. To showcase the popular preprint destination venues, we present twenty-four top journals and fourteen top conferences with the highest numbers of preprint papers in Appendix \ref{appendix:jc}. Four top journals from each domain are listed. Since there are not many disciplines which regard peer-reviewed conferences as primary publishing venues except for \textit{computer science},  the top conferences listed are all in \textit{computer science} domain.


\subsection{Preprint for disease outbreak response}
In September 2015, a statement from a WHO Consultation urged a timely share of pre-publication data during public health emergencies~\cite{who2015}. In the following year, Wellcome Trust, together with 30 other global health bodies, called for rapid and open access for Zika research data \cite{welcom2016}. Both statements emphasized the importance and necessity of sharing preliminary data faster in such disease outbreaks. The COVID-19 pandemic that started in December 2019 has infected more than 34.8 million people and caused 1 million deaths around the globe by October 2020 according to the WHO situation reports~\cite{who2020}. Compared with other outbreaks in the 21st century, such as SARS, H1N1 swine flu, MERS, Ebola, and Zika, scientists respond to COVID-19 much faster this time. The number of papers related to COVID-19, especially preprint papers, has surpassed any other outbreak topics. Preprint has been unprecedentedly utilized by researchers to share their studies on pathogenesis, epidemiological characteristic and estimation, treatment, vaccine, etc. Fraser et al.~\cite{Fraser2020} reported that 37.5\% of COVID-19 papers are hosted on preprint servers and the weekly new COVID-19 preprints is as many as 250. The top preprint articles have been reported in more than 300 news and tweeted over 10 thousand times. The swift release process of preprints not only benefits researchers to access the latest findings earlier but also bridges the communication between laboratories and the public more efficiently at this unique challenging time. 

In conclusion, preprints have implemented several screening processes for the quality control. 41\% of preprints are published in peer reviewed journals or conferences, and the destination venues are as influential as  papers without a preprint version, if not more. Preprint contributes to the research community by facilitating rapid open access of research results which is highly valuable during public health emergencies.

\section{Related Work}
To the best of our knowledge, there are more than sixty online preprint servers active today. Despite the absence of an overview which covers all disciplines with large scale data, various evaluations have been carried out with a specific subject area or server. The most frequently evaluated preprint servers are arXiv and bioRxiv; popular evaluated subject areas are high energy physics, biomedical, economics, and COVID-19. Gentil-Beccot et al.~\cite{Beccot2009} analyzed the citation and reading behavior in arXiv with 286,180 high energy physics journal articles and preprints which appeared from 1991 to 2007. Kelk and Devine~\cite{Kelk2012} did scienceographic comparison between arXiv and viXra by sampling 20 physics papers appeared between 2007 and 2012 from each of the two servers. A detailed overview of all 37,648 preprints uploaded to bioRxiv from 2013 to 2018 were performed by Abdill and Belkham~\cite{Abdill2019}. Serghio et al.~\cite{Serghio2018} assessed 776 bioRxiv preprints posted between 2013 and 2017 to measure their Altmetric scores and citations. A similar work assessed 5,405 bioRxiv preprint was done by Fu and Hughey~\cite{Fu2019}. Baumann and Wohlrabe~\cite{Baumann2020} investigated four working paper series in RePEc to estimate the publish rate of economics working papers. Li et al.~\cite{Li2015} compared the citations received between four repositories: arXiv, RePEc, SSRN and PMC by sampling 384 papers. While the COVID-19 pandemic is affecting the whole world, vast number of studies related to this disease were carried out, including analysis on the preprints' performance during the pandemic~\cite{Fraser2020,Gianola2020,nb2020,Kwon2020}. Key characteristics and policies of 44 biomedical preprint platforms were reviewed in \cite{Kirkham2020}. Besides data analysis, there were also many discussions related to preprints, such as the reasons to use and not to use preprints and the value to early career researchers~\cite{Sarabipour2019, Kaiser2017,Mayo2020,Bourne2017}.

\section{Conclusion}
In this work, we provide a quantitative overview of preprints from all disciplines in the past thirty years. Despite the increasing number of preprint servers and effort from major research funders adopting policies to promote preprints, the preprint has not been a popular form for most scientists. Generally \textit{physics} and \textit{mathematics} are the two disciplines that adopt preprint culture the best. They have long history of using online preprint archives, produce the most number of preprints (50\% of all preprints), and have high preprints to all-paper ratios  as well as highest preprint publication rates. \textit{Biology} has significant increase in recent years, both in the preprints size and the publication rate. 

One major advantage of preprint is its immediate online release in contrast to the long peer-reviewed publishing process. This study shows that by using preprints, individual researchers communicate their findings to the audiences 14 months ahead of peer-reviewed papers, and the work is cited five times more compare to the articles only published in peer-reviewed venues without a preprint version. Researchers in \textit{economics}  gain the most time and citation advantage when they utilize preprints.

Besides the benefit to individual researchers, preprints also contribute to the research community by providing a platform which shares valuable research results in a timely fashion. The quality of  preprint papers is assessed through the discussion of existing quality control strategies, the preprint publication rate, and the quality of their destination venues. The 41\% publication rate and the similar destination impact factor distribution indicate the quality of preprints is on par with peer-reviewed papers. Last but not least, we emphasize the important role that preprint plays in response to disease outbreaks, and its swift publishing mechanism shows unbeatable advantage in such time sensitive events.

In the past thirty years, our analysis has shown that the format of scholarly communication has evolved together with the rapid Web technological development. The scientific publishing landscape has been gradually shifted to embrace the preprint culture. The ultimate goal of  scholarly communication is to disseminate the knowledge and new discoveries more rapidly, widely, and cost-effectively. Our work provides quantitative evidence to demonstrate that online preprint services could make us one step closer to this grand goal. 
We have witnessed the fast adoption and flourish of the preprint culture in biomedical domains in the past five years with the promotional efforts from research organizations, publishers, and high profile researchers. Despite our extensive analysis of the positive impacts of preprints, there are still uncharacterized facets of the reserved attitudes towards preprints among many researchers. How to better promote preprints in different communities will need a closer look by domain experts and to integrate other data besides scholarly publications themselves.

\bibliography{main}


\begin{thebibliography}{49}

%%% the Web address.

\ifx \showCODEN    \undefined \def \showCODEN     #1{\unskip}     \fi
\ifx \showDOI      \undefined \def \showDOI       #1{#1}\fi
\ifx \showISBNx    \undefined \def \showISBNx     #1{\unskip}     \fi
\ifx \showISBNxiii \undefined \def \showISBNxiii  #1{\unskip}     \fi
\ifx \showISSN     \undefined \def \showISSN      #1{\unskip}     \fi
\ifx \showLCCN     \undefined \def \showLCCN      #1{\unskip}     \fi
\ifx \shownote     \undefined \def \shownote      #1{#1}          \fi
\ifx \showarticletitle \undefined \def \showarticletitle #1{#1}   \fi
\ifx \showURL      \undefined \def \showURL       {\relax}        \fi
% invisible to TeX
\providecommand\bibfield[2]{#2}
\providecommand\bibinfo[2]{#2}
\providecommand\natexlab[1]{#1}
\providecommand\showeprint[2][]{arXiv:#2}

\bibitem[\protect\citeauthoryear{Abdill and Blekhman}{Abdill and
  Blekhman}{2019}]%
        {Abdill2019}
\bibfield{author}{\bibinfo{person}{Richard~J. Abdill} {and}
  \bibinfo{person}{Ran Blekhman}.} \bibinfo{year}{2019}\natexlab{}.
\newblock \showarticletitle{Tracking the popularity and outcomes of all bioRxiv
  preprints}.
\newblock \bibinfo{journal}{\emph{eLife}}  \bibinfo{volume}{8}
  (\bibinfo{date}{April} \bibinfo{year}{2019}).
\newblock
\urldef\tempurl%
\url{https://doi.org/10.7554/eLife.45133}
\showDOI{\tempurl}


\bibitem[\protect\citeauthoryear{Alberts, Hanson, and Kelner}{Alberts
  et~al\mbox{.}}{2008}]%
        {Alberts2008}
\bibfield{author}{\bibinfo{person}{Bruce Alberts}, \bibinfo{person}{Brooks
  Hanson}, {and} \bibinfo{person}{Katrina~L. Kelner}.}
  \bibinfo{year}{2008}\natexlab{}.
\newblock \showarticletitle{Reviewing peer review}.
\newblock \bibinfo{journal}{\emph{Science}}  \bibinfo{volume}{321}
  (\bibinfo{date}{July} \bibinfo{year}{2008}), \bibinfo{pages}{15}.
\newblock
\urldef\tempurl%
\url{https://doi.org/10.1126/science.1162115}
\showDOI{\tempurl}


\bibitem[\protect\citeauthoryear{Allison, Brown, George, and Kaiser}{Allison
  et~al\mbox{.}}{2016}]%
        {Allison2016}
\bibfield{author}{\bibinfo{person}{David~B. Allison}, \bibinfo{person}{Andrew~W
  Brown}, \bibinfo{person}{Brandon~J George}, {and} \bibinfo{person}{Kathryn~A
  Kaiser}.} \bibinfo{year}{2016}\natexlab{}.
\newblock \showarticletitle{Reproducibility: A tragedy of errors}.
\newblock \bibinfo{journal}{\emph{Nature}}  \bibinfo{volume}{530}
  (\bibinfo{date}{Feb.} \bibinfo{year}{2016}), \bibinfo{pages}{27--29}.
\newblock
\urldef\tempurl%
\url{https://doi.org/10.1038/530027a}
\showDOI{\tempurl}


\bibitem[\protect\citeauthoryear{arXiv}{arXiv}{2019}]%
        {ArxivSub}
\bibfield{author}{\bibinfo{person}{arXiv}.} \bibinfo{year}{2019}\natexlab{}.
\newblock \bibinfo{title}{Our Moderation Process}.
\newblock
\newblock
\urldef\tempurl%
\url{http://blog.arxiv.org/2019/08/29/our-moderation-process/}
\showURL{%
Retrieved September 30, 2020 from \tempurl}


\bibitem[\protect\citeauthoryear{arXiv}{arXiv}{2020}]%
        {ArxivEndor}
\bibfield{author}{\bibinfo{person}{arXiv}.} \bibinfo{year}{2020}\natexlab{}.
\newblock \bibinfo{title}{The arXiv endorsement system}.
\newblock
\newblock
\urldef\tempurl%
\url{https://arxiv.org/help/endorsement}
\showURL{%
Retrieved August 31, 2020 from \tempurl}


\bibitem[\protect\citeauthoryear{Balietti}{Balietti}{2016}]%
        {Balietti2016}
\bibfield{author}{\bibinfo{person}{Stefano Balietti}.}
  \bibinfo{year}{2016}\natexlab{}.
\newblock \showarticletitle{Here’s how competition makes peer review more
  unfair}.
\newblock \bibinfo{journal}{\emph{The conversation}} (\bibinfo{date}{Aug.}
  \bibinfo{year}{2016}).
\newblock
\urldef\tempurl%
\url{https://theconversation.com/heres-how-competition-makes-peer-review-more-unfair-62936}
\showURL{%
\tempurl}


\bibitem[\protect\citeauthoryear{Baumann and Wohlrabe}{Baumann and
  Wohlrabe}{2020}]%
        {Baumann2020}
\bibfield{author}{\bibinfo{person}{Alexandra Baumann} {and}
  \bibinfo{person}{Klaus Wohlrabe}.} \bibinfo{year}{2020}\natexlab{}.
\newblock \showarticletitle{Where have all the working papers gone? Evidence
  from four major economics working paper series}.
\newblock \bibinfo{journal}{\emph{Scientometrics}} (\bibinfo{date}{July}
  \bibinfo{year}{2020}), \bibinfo{pages}{2433--2411}.
\newblock
\urldef\tempurl%
\url{https://doi.org/10.1007/s11192-020-03570-x}
\showDOI{\tempurl}


\bibitem[\protect\citeauthoryear{Biotechnology}{Biotechnology}{2020}]%
        {nb2020}
\bibfield{editor}{\bibinfo{person}{Nature Biotechnology}} (Ed.).
  \bibinfo{year}{2020}\natexlab{}.
\newblock \showarticletitle{All that’s fit to preprint}.
\newblock \bibinfo{journal}{\emph{Nature Biotechnology}} (\bibinfo{date}{May}
  \bibinfo{year}{2020}).
\newblock
\urldef\tempurl%
\url{https://doi.org/10.1038/s41587-020-0536-x}
\showDOI{\tempurl}


\bibitem[\protect\citeauthoryear{Bourne, Polka, Vale, and Kiley}{Bourne
  et~al\mbox{.}}{2017}]%
        {Bourne2017}
\bibfield{author}{\bibinfo{person}{Philip~E. Bourne},
  \bibinfo{person}{Jessica~K. Polka}, \bibinfo{person}{Ronald~D. Vale}, {and}
  \bibinfo{person}{Robert Kiley}.} \bibinfo{year}{2017}\natexlab{}.
\newblock \showarticletitle{Ten simple rules to consider regarding preprint
  submission}.
\newblock \bibinfo{journal}{\emph{PLoS Comput Biol}} \bibinfo{volume}{13},
  \bibinfo{number}{5} (\bibinfo{date}{May} \bibinfo{year}{2017}).
\newblock
\urldef\tempurl%
\url{https://doi.org/10.1371/journal.pcbi.1005473}
\showDOI{\tempurl}


\bibitem[\protect\citeauthoryear{Clemens}{Clemens}{2020}]%
        {Clemens}
\bibfield{author}{\bibinfo{person}{Anna Clemens}.}
  \bibinfo{year}{2020}\natexlab{}.
\newblock \bibinfo{title}{5 common concerns about publishing preprints}.
\newblock
\newblock
\urldef\tempurl%
\url{https://www.annaclemens.com/blog/downsides-publishing-preprint}
\showURL{%
Retrieved September 15, 2020 from \tempurl}


\bibitem[\protect\citeauthoryear{Fortunato, Bergstrom, Börner, Evans, Helbing,
  Milojević, Petersen, Radicchi, Sinatra, Uzzi, Vespignani, Waltman, Wang, and
  Barabási}{Fortunato et~al\mbox{.}}{2018}]%
        {Santo2018}
\bibfield{author}{\bibinfo{person}{Santo Fortunato}, \bibinfo{person}{Carl~T.
  Bergstrom}, \bibinfo{person}{Katy Börner}, \bibinfo{person}{James~A. Evans},
  \bibinfo{person}{Dirk Helbing}, \bibinfo{person}{Staša Milojević},
  \bibinfo{person}{Alexander~M. Petersen}, \bibinfo{person}{Filippo Radicchi},
  \bibinfo{person}{Roberta Sinatra}, \bibinfo{person}{Brian Uzzi},
  \bibinfo{person}{Alessandro Vespignani}, \bibinfo{person}{Ludo Waltman},
  \bibinfo{person}{Dashun Wang}, {and} \bibinfo{person}{Albert-László
  Barabási}.} \bibinfo{year}{2018}\natexlab{}.
\newblock \showarticletitle{Science of science}.
\newblock \bibinfo{journal}{\emph{Science}} (\bibinfo{date}{March}
  \bibinfo{year}{2018}).
\newblock
\urldef\tempurl%
\url{https://science.sciencemag.org/content/359/6379/eaao0185.full}
\showURL{%
\tempurl}


\bibitem[\protect\citeauthoryear{Fountain}{Fountain}{2014}]%
        {Fountain2014}
\bibfield{author}{\bibinfo{person}{Henry Fountain}.}
  \bibinfo{year}{2014}\natexlab{}.
\newblock \showarticletitle{Science Journal Pulls 60 Papers in Peer-Review
  Fraud}.
\newblock \bibinfo{journal}{\emph{The New York Times}} (\bibinfo{date}{July}
  \bibinfo{year}{2014}), \bibinfo{pages}{3}.
\newblock
\urldef\tempurl%
\url{https://www.nytimes.com/2014/07/11/science/science-journal-pulls-60-papers-in-peer-review-fraud.html}
\showURL{%
\tempurl}


\bibitem[\protect\citeauthoryear{Fraser, Brierley, Dey, Polka, Pálfy, and
  Coates}{Fraser et~al\mbox{.}}{2020}]%
        {Fraser2020}
\bibfield{author}{\bibinfo{person}{Nicholas Fraser}, \bibinfo{person}{Liam
  Brierley}, \bibinfo{person}{Gautam Dey}, \bibinfo{person}{Jessica~K Polka},
  \bibinfo{person}{Máté Pálfy}, {and} \bibinfo{person}{Jonathon~Alexis
  Coates}.} \bibinfo{year}{2020}\natexlab{}.
\newblock \showarticletitle{Preprinting a pandemic: the role of preprints in
  the COVID-19 pandemic}.
\newblock \bibinfo{journal}{\emph{bioRxiv}} (\bibinfo{date}{May}
  \bibinfo{year}{2020}).
\newblock
\urldef\tempurl%
\url{https://doi.org/10.1101/2020.05.22.111294}
\showDOI{\tempurl}


\bibitem[\protect\citeauthoryear{Fu and Hughey}{Fu and Hughey}{2019}]%
        {Fu2019}
\bibfield{author}{\bibinfo{person}{Darwin~Y Fu} {and} \bibinfo{person}{Jacob~J
  Hughey}.} \bibinfo{year}{2019}\natexlab{}.
\newblock \showarticletitle{Releasing a preprint is associated with more
  attention and citations for the peer-reviewed article}.
\newblock \bibinfo{journal}{\emph{eLife}} (\bibinfo{date}{Dec.}
  \bibinfo{year}{2019}).
\newblock
\urldef\tempurl%
\url{https://doi.org/10.7554/eLife.52646}
\showDOI{\tempurl}


\bibitem[\protect\citeauthoryear{Garisto}{Garisto}{2019}]%
        {Garisto19}
\bibfield{author}{\bibinfo{person}{Daniel Garisto}.}
  \bibinfo{year}{2019}\natexlab{}.
\newblock \showarticletitle{Preprints Make Inroads Outside of Physics}.
\newblock \bibinfo{journal}{\emph{APS News}} \bibinfo{volume}{28},
  \bibinfo{number}{9} (\bibinfo{date}{Oct.} \bibinfo{year}{2019}).
\newblock
\urldef\tempurl%
\url{https://www.aps.org/publications/apsnews/201909/preprints.cfm}
\showURL{%
\tempurl}


\bibitem[\protect\citeauthoryear{Gentil-Beccot, Mele, and Brooks}{Gentil-Beccot
  et~al\mbox{.}}{2010}]%
        {Beccot2009}
\bibfield{author}{\bibinfo{person}{Anne Gentil-Beccot},
  \bibinfo{person}{Salvatore Mele}, {and} \bibinfo{person}{Travis~C. Brooks}.}
  \bibinfo{year}{2010}\natexlab{}.
\newblock \showarticletitle{Citing and Reading Behaviours in High-Energy
  Physics}.
\newblock \bibinfo{journal}{\emph{Scientometrics}}  \bibinfo{volume}{84}
  (\bibinfo{date}{Dec.} \bibinfo{year}{2010}), \bibinfo{pages}{345--355}.
\newblock
\urldef\tempurl%
\url{https://doi.org/10.1007/s11192-009-0111-1}
\showDOI{\tempurl}


\bibitem[\protect\citeauthoryear{Gianola, Jesus, Bargeri, and
  Castellini}{Gianola et~al\mbox{.}}{2020}]%
        {Gianola2020}
\bibfield{author}{\bibinfo{person}{Silvia Gianola}, \bibinfo{person}{Tiago~S
  Jesus}, \bibinfo{person}{Silvia Bargeri}, {and} \bibinfo{person}{Greta
  Castellini}.} \bibinfo{year}{2020}\natexlab{}.
\newblock \showarticletitle{Characteristics of academic publications,
  preprints, and registered clinical trials on the COVID-19 pandemic}.
\newblock \bibinfo{journal}{\emph{PLoS One}} \bibinfo{volume}{6},
  \bibinfo{number}{15} (\bibinfo{date}{Oct.} \bibinfo{year}{2020}).
\newblock
\urldef\tempurl%
\url{https://doi.org/10.1371/journal.pone.0240123}
\showDOI{\tempurl}


\bibitem[\protect\citeauthoryear{Ginsparg}{Ginsparg}{2011}]%
        {Ginsparg2011}
\bibfield{author}{\bibinfo{person}{Paul Ginsparg}.}
  \bibinfo{year}{2011}\natexlab{}.
\newblock \showarticletitle{ArXiv at 20}.
\newblock \bibinfo{journal}{\emph{Nature}}  \bibinfo{volume}{476}
  (\bibinfo{date}{Aug.} \bibinfo{year}{2011}), \bibinfo{pages}{145--147}.
\newblock
\urldef\tempurl%
\url{https://doi.org/10.1038/476145a}
\showDOI{\tempurl}


\bibitem[\protect\citeauthoryear{Golosovsky and Solomon}{Golosovsky and
  Solomon}{2012}]%
        {Golo2012}
\bibfield{author}{\bibinfo{person}{Michael Golosovsky} {and}
  \bibinfo{person}{Sorin Solomon}.} \bibinfo{year}{2012}\natexlab{}.
\newblock \showarticletitle{Runaway events dominate the heavy tail of citation
  distributions}.
\newblock \bibinfo{journal}{\emph{European Physical Journal-special Topics}}
  \bibinfo{volume}{205} (\bibinfo{year}{2012}), \bibinfo{pages}{303--311}.
\newblock


\bibitem[\protect\citeauthoryear{IMF}{IMF}{2020}]%
        {IMF}
\bibfield{author}{\bibinfo{person}{IMF}.} \bibinfo{year}{2020}\natexlab{}.
\newblock \bibinfo{title}{International Monetary Fund}.
\newblock
\newblock
\urldef\tempurl%
\url{https://www.imf.org/en/Publications/Search?series=IMF\%20Working\%20Papers}
\showURL{%
Retrieved August 31, 2020 from \tempurl}


\bibitem[\protect\citeauthoryear{Johnson and Chiarelli}{Johnson and
  Chiarelli}{2019}]%
        {Johnson2019}
\bibfield{author}{\bibinfo{person}{Rob Johnson} {and} \bibinfo{person}{Andrea
  Chiarelli}.} \bibinfo{year}{2019}\natexlab{}.
\newblock \showarticletitle{The Second Wave of Preprint Servers: How Can
  Publishers Keep Afloat?}
\newblock \bibinfo{journal}{\emph{The scholarly kitchen}} (\bibinfo{date}{Oct.}
  \bibinfo{year}{2019}).
\newblock
\urldef\tempurl%
\url{https://scholarlykitchen.sspnet.org/2019/10/16/the-second-wave-of-preprint-servers-how-can-publishers-keep-afloat/}
\showURL{%
\tempurl}


\bibitem[\protect\citeauthoryear{Kaiser}{Kaiser}{2017}]%
        {Kaiser2017}
\bibfield{author}{\bibinfo{person}{Jocelyn Kaiser}.}
  \bibinfo{year}{2017}\natexlab{}.
\newblock \showarticletitle{Are preprints the future of biology? A survival
  guide for scientists}.
\newblock \bibinfo{journal}{\emph{Science News}} (\bibinfo{date}{Sept.}
  \bibinfo{year}{2017}).
\newblock
\urldef\tempurl%
\url{https://doi.org/0.1126/science.aaq0747}
\showDOI{\tempurl}


\bibitem[\protect\citeauthoryear{Kelk and Devine}{Kelk and Devine}{2012}]%
        {Kelk2012}
\bibfield{author}{\bibinfo{person}{David Kelk} {and} \bibinfo{person}{David
  Devine}.} \bibinfo{year}{2012}\natexlab{}.
\newblock \showarticletitle{A Scienceographic Comparison of Physics Papers from
  the arXiv and viXra Archives}.
\newblock \bibinfo{journal}{\emph{arXiv}} (\bibinfo{date}{Nov.}
  \bibinfo{year}{2012}).
\newblock
\urldef\tempurl%
\url{https://doi.org/1211.1036}
\showDOI{\tempurl}


\bibitem[\protect\citeauthoryear{Kirkham, Penfold, Murphy, Boutron, Ioannidis,
  Polka, and Moher}{Kirkham et~al\mbox{.}}{2020}]%
        {Kirkham2020}
\bibfield{author}{\bibinfo{person}{Jamie~J Kirkham}, \bibinfo{person}{Naomi
  Penfold}, \bibinfo{person}{Fiona Murphy}, \bibinfo{person}{Isabelle Boutron},
  \bibinfo{person}{John~PA Ioannidis}, \bibinfo{person}{Jessica~K Polka}, {and}
  \bibinfo{person}{David Moher}.} \bibinfo{year}{2020}\natexlab{}.
\newblock \showarticletitle{A systematic examination of preprint platforms for
  use in the medical and biomedical sciences setting}.
\newblock \bibinfo{journal}{\emph{bioRxiv}} (\bibinfo{year}{2020}).
\newblock
\urldef\tempurl%
\url{https://doi.org/10.1101/2020.04.27.063578}
\showDOI{\tempurl}


\bibitem[\protect\citeauthoryear{Kwon}{Kwon}{2020}]%
        {Kwon2020}
\bibfield{author}{\bibinfo{person}{Diana Kwon}.}
  \bibinfo{year}{2020}\natexlab{}.
\newblock \showarticletitle{How swamped preprint servers are blocking bad
  coronavirus research}.
\newblock \bibinfo{journal}{\emph{Nature}}  \bibinfo{volume}{581}
  (\bibinfo{date}{May} \bibinfo{year}{2020}), \bibinfo{pages}{130--131}.
\newblock
\urldef\tempurl%
\url{https://doi.org/10.1038/d41586-020-01394-6}
\showDOI{\tempurl}


\bibitem[\protect\citeauthoryear{Laboratory}{Laboratory}{2020a}]%
        {biorxivSub}
\bibfield{author}{\bibinfo{person}{Cold Spring~Harbor Laboratory}.}
  \bibinfo{year}{2020}\natexlab{a}.
\newblock \bibinfo{title}{Submission Guide}.
\newblock
\newblock
\urldef\tempurl%
\url{https://www.biorxiv.org/submit-a-manuscript}
\showURL{%
Retrieved August 31, 2020 from \tempurl}


\bibitem[\protect\citeauthoryear{Laboratory}{Laboratory}{2020b}]%
        {medrxivSub}
\bibfield{author}{\bibinfo{person}{Cold Spring~Harbor Laboratory}.}
  \bibinfo{year}{2020}\natexlab{b}.
\newblock \bibinfo{title}{Submission Guide}.
\newblock
\newblock
\urldef\tempurl%
\url{https://www.medrxiv.org/submit-a-manuscript}
\showURL{%
Retrieved August 31, 2020 from \tempurl}


\bibitem[\protect\citeauthoryear{Li, Thelwall, and Kousha}{Li
  et~al\mbox{.}}{2015}]%
        {Li2015}
\bibfield{author}{\bibinfo{person}{Xuemei Li}, \bibinfo{person}{Mike Thelwall},
  {and} \bibinfo{person}{Kayvan Kousha}.} \bibinfo{year}{2015}\natexlab{}.
\newblock \showarticletitle{The role of arXiv, RePEc, SSRN and PMC in formal
  scholarly communication}.
\newblock \bibinfo{journal}{\emph{Aslib Journal of Information Management}}
  \bibinfo{volume}{67}, \bibinfo{number}{6} (\bibinfo{date}{Nov.}
  \bibinfo{year}{2015}), \bibinfo{pages}{614--635}.
\newblock
\urldef\tempurl%
\url{https://doi.org/10.1108/AJIM-03-2015-0049}
\showDOI{\tempurl}


\bibitem[\protect\citeauthoryear{Mayo-Yánez}{Mayo-Yánez}{2020}]%
        {Mayo2020}
\bibfield{author}{\bibinfo{person}{Miguel Mayo-Yánez}.}
  \bibinfo{year}{2020}\natexlab{}.
\newblock \showarticletitle{Research during SARS-CoV-2 pandemic: To
  “Preprint” or not to “Preprint”, that is the question}.
\newblock \bibinfo{journal}{\emph{Med Clin (Barc)}} \bibinfo{volume}{2},
  \bibinfo{number}{155} (\bibinfo{date}{July} \bibinfo{year}{2020}),
  \bibinfo{pages}{86--87}.
\newblock
\urldef\tempurl%
\url{https://doi.org/10.1016/j.medcli.2020.05.002}
\showDOI{\tempurl}


\bibitem[\protect\citeauthoryear{Moore}{Moore}{1965}]%
        {Moore1965}
\bibfield{author}{\bibinfo{person}{C.~Alan Moore}.}
  \bibinfo{year}{1965}\natexlab{}.
\newblock \showarticletitle{Preprints. An Old Information Device with New
  Outlooks}.
\newblock \bibinfo{journal}{\emph{J. Chem. Doc.}} \bibinfo{volume}{5},
  \bibinfo{number}{3} (\bibinfo{date}{Aug.} \bibinfo{year}{1965}),
  \bibinfo{pages}{126--128}.
\newblock
\urldef\tempurl%
\url{https://doi.org/10.1021/c160018a003}
\showDOI{\tempurl}


\bibitem[\protect\citeauthoryear{Noorden}{Noorden}{2016}]%
        {Noorden2016}
\bibfield{author}{\bibinfo{person}{Richard~Van Noorden}.}
  \bibinfo{year}{2016}\natexlab{}.
\newblock \showarticletitle{Social-sciences preprint server snapped up by
  publishing giant Elsevier}.
\newblock \bibinfo{journal}{\emph{Nature}} (\bibinfo{date}{May}
  \bibinfo{year}{2016}).
\newblock
\urldef\tempurl%
\url{https://doi.org/doi:10.1038/nature.2016.19932}
\showDOI{\tempurl}


\bibitem[\protect\citeauthoryear{of~Applied~Mathematics}{of~Applied~Mathematics}{2020}]%
        {Keldysh}
\bibfield{author}{\bibinfo{person}{Keldysh~Institute of Applied~Mathematics}.}
  \bibinfo{year}{2020}\natexlab{}.
\newblock \bibinfo{title}{Keldysh Institute Preprints}.
\newblock
\newblock
\urldef\tempurl%
\url{https://library.keldysh.ru/preprints/default.asp?lg=e}
\showURL{%
Retrieved August 31, 2020 from \tempurl}


\bibitem[\protect\citeauthoryear{Organization}{Organization}{2015}]%
        {who2015}
\bibfield{author}{\bibinfo{person}{World~Health Organization}.}
  \bibinfo{year}{2015}\natexlab{}.
\newblock \bibinfo{title}{Developing global norms for sharing data and results
  during public health emergencies}.
\newblock
\newblock
\urldef\tempurl%
\url{https://www.who.int/medicines/ebola-treatment/blueprint_phe_data-share-results/en/}
\showURL{%
Retrieved Octorber 11, 2020 from \tempurl}


\bibitem[\protect\citeauthoryear{Organization}{Organization}{2020}]%
        {who2020}
\bibfield{author}{\bibinfo{person}{World~Health Organization}.}
  \bibinfo{year}{2020}\natexlab{}.
\newblock \bibinfo{title}{Weekly epidemiological updates coronavirus disease
  (COVID-19)}.
\newblock
\newblock
\urldef\tempurl%
\url{https://www.who.int/docs/default-source/coronaviruse/situation-reports/20201005-weekly-epi-update-8.pdf}
\showURL{%
Retrieved Octorber 11, 2020 from \tempurl}


\bibitem[\protect\citeauthoryear{Ozler}{Ozler}{2011}]%
        {Ozler2011}
\bibfield{author}{\bibinfo{person}{Berk Ozler}.}
  \bibinfo{year}{2011}\natexlab{}.
\newblock \bibinfo{title}{Working Papers are NOT Working.}
\newblock
\newblock
\urldef\tempurl%
\url{https://blogs.worldbank.org/impactevaluations/working-papers-are-not-working}
\showURL{%
Retrieved Octorber 11, 2020 from \tempurl}


\bibitem[\protect\citeauthoryear{RePEc}{RePEc}{2020}]%
        {RePEc}
\bibfield{author}{\bibinfo{person}{RePEc}.} \bibinfo{year}{2020}\natexlab{}.
\newblock \bibinfo{title}{RePEc History}.
\newblock
\newblock
\urldef\tempurl%
\url{https://ideas.repec.org/history.html}
\showURL{%
Retrieved August 31, 2020 from \tempurl}


\bibitem[\protect\citeauthoryear{Rosenfeld, Wakerling, Addis, Gex, and
  Taylor}{Rosenfeld et~al\mbox{.}}{1970}]%
        {Rosen1970}
\bibfield{author}{\bibinfo{person}{A. Rosenfeld}, \bibinfo{person}{R.K.
  Wakerling}, \bibinfo{person}{L. Addis}, \bibinfo{person}{R. Gex}, {and}
  \bibinfo{person}{R.J. Taylor}.} \bibinfo{year}{1970}\natexlab{}.
\newblock \showarticletitle{Preprints in particles and fields}.
\newblock \bibinfo{journal}{\emph{SLAC-PUB-0710}} (\bibinfo{date}{Feb.}
  \bibinfo{year}{1970}).
\newblock
\urldef\tempurl%
\url{https://www.slac.stanford.edu/cgi-bin/getdoc/slac-pub-0710.pdf}
\showURL{%
\tempurl}


\bibitem[\protect\citeauthoryear{Sarabipour, Debat, Emmott, Burgess,
  Schwessinger, and Hensel}{Sarabipour et~al\mbox{.}}{2019}]%
        {Sarabipour2019}
\bibfield{author}{\bibinfo{person}{Sarvenaz Sarabipour},
  \bibinfo{person}{Humberto~J. Debat}, \bibinfo{person}{Edward Emmott},
  \bibinfo{person}{Steven~J. Burgess}, \bibinfo{person}{Benjamin Schwessinger},
  {and} \bibinfo{person}{Zach Hensel}.} \bibinfo{year}{2019}\natexlab{}.
\newblock \showarticletitle{On the value of preprints: An early career
  researcher perspective}.
\newblock \bibinfo{journal}{\emph{PLoS Biology}} (\bibinfo{date}{Feb.}
  \bibinfo{year}{2019}).
\newblock
\urldef\tempurl%
\url{https://doi.org/10.1371/journal.pbio.3000151}
\showDOI{\tempurl}


\bibitem[\protect\citeauthoryear{Serghiou, MBChB(Hons), and Ioannidis}{Serghiou
  et~al\mbox{.}}{2018}]%
        {Serghio2018}
\bibfield{author}{\bibinfo{person}{Stylianos Serghiou},
  \bibinfo{person}{MBChB(Hons)}, {and} \bibinfo{person}{John~P.A. Ioannidis}.}
  \bibinfo{year}{2018}\natexlab{}.
\newblock \showarticletitle{Altmetric Scores, Citations, and Publication of
  Studies Posted as Preprints}.
\newblock \bibinfo{journal}{\emph{JAMA}} \bibinfo{volume}{139},
  \bibinfo{number}{4} (\bibinfo{date}{Jan.} \bibinfo{year}{2018}),
  \bibinfo{pages}{402--404}.
\newblock
\urldef\tempurl%
\url{https://doi.org/10.1001/jama.2017.21168}
\showDOI{\tempurl}


\bibitem[\protect\citeauthoryear{{Shen}, {Ma}, and {Wang}}{{Shen}
  et~al\mbox{.}}{2018}]%
        {shen2018a}
\bibfield{author}{\bibinfo{person}{Zhihong {Shen}}, \bibinfo{person}{Hao {Ma}},
  {and} \bibinfo{person}{Kuansan {Wang}}.} \bibinfo{year}{2018}\natexlab{}.
\newblock \showarticletitle{A web-scale system for scientific knowledge
  exploration}. In \bibinfo{booktitle}{\emph{ACL 2018: 56th Annual Meeting of
  the Association for Computational Linguistics}}. \bibinfo{pages}{87--92}.
\newblock


\bibitem[\protect\citeauthoryear{Sinha, Shen, Song, Ma, Eide, Hsu, and
  Wang}{Sinha et~al\mbox{.}}{2015}]%
        {Sinha-15}
\bibfield{author}{\bibinfo{person}{Arnab Sinha}, \bibinfo{person}{Zhihong
  Shen}, \bibinfo{person}{Yang Song}, \bibinfo{person}{Hao Ma},
  \bibinfo{person}{Darrin Eide}, \bibinfo{person}{Bo-June~(Paul) Hsu}, {and}
  \bibinfo{person}{Kuangsan Wang}.} \bibinfo{year}{2015}\natexlab{}.
\newblock \showarticletitle{An Overview of Microsoft Academic Service (MA) and
  Applications}. In \bibinfo{booktitle}{\emph{Proceedings of the 24th
  International Conference on World Wide Web (WWW '15 Companion)}}.
  \bibinfo{publisher}{ACM}, \bibinfo{address}{New York, NY, USA},
  \bibinfo{pages}{243--246}.
\newblock
\urldef\tempurl%
\url{https://doi.org/10.1145/2740908.2742839}
\showDOI{\tempurl}


\bibitem[\protect\citeauthoryear{Smith}{Smith}{2006}]%
        {Smith2006}
\bibfield{author}{\bibinfo{person}{Richard Smith}.}
  \bibinfo{year}{2006}\natexlab{}.
\newblock \showarticletitle{Peer review: a flawed process at the heart of
  science and journals}.
\newblock \bibinfo{journal}{\emph{Journal of the Royal Society of Medicine}}
  \bibinfo{volume}{99}, \bibinfo{number}{4} (\bibinfo{date}{April}
  \bibinfo{year}{2006}), \bibinfo{pages}{178--182}.
\newblock
\urldef\tempurl%
\url{https://doi.org/10.1258/jrsm.99.4.178}
\showDOI{\tempurl}


\bibitem[\protect\citeauthoryear{SSRN}{SSRN}{2020a}]%
        {SSRNStats}
\bibfield{author}{\bibinfo{person}{SSRN}.} \bibinfo{year}{2020}\natexlab{a}.
\newblock \bibinfo{title}{SSRN eLibrary Statistics}.
\newblock
\newblock
\urldef\tempurl%
\url{https://papers.ssrn.com/sol3/DisplayAbstractSearch.cfm}
\showURL{%
Retrieved September 11, 2020 from \tempurl}


\bibitem[\protect\citeauthoryear{SSRN}{SSRN}{2020b}]%
        {SSRNSub}
\bibfield{author}{\bibinfo{person}{SSRN}.} \bibinfo{year}{2020}\natexlab{b}.
\newblock \bibinfo{title}{SSRN FAQ Page}.
\newblock
\newblock
\urldef\tempurl%
\url{https://www.ssrn.com/index.cfm/en/ssrn-faq/}
\showURL{%
Retrieved August 31, 2020 from \tempurl}


\bibitem[\protect\citeauthoryear{Stahel and Moore}{Stahel and Moore}{2014}]%
        {Stahel2014}
\bibfield{author}{\bibinfo{person}{Philip~F Stahel} {and}
  \bibinfo{person}{Ernest~E Moore}.} \bibinfo{year}{2014}\natexlab{}.
\newblock \showarticletitle{Peer review for biomedical publications: we can
  improve the system}.
\newblock \bibinfo{journal}{\emph{BMC Med.}} \bibinfo{volume}{12},
  \bibinfo{number}{179} (\bibinfo{date}{Sept.} \bibinfo{year}{2014}).
\newblock
\urldef\tempurl%
\url{https://doi.org/10.1186/s12916-014-0179-1}
\showDOI{\tempurl}


\bibitem[\protect\citeauthoryear{Stegehuis, Litvak, and Waltman}{Stegehuis
  et~al\mbox{.}}{2012}]%
        {stege2012}
\bibfield{author}{\bibinfo{person}{C~Clara Stegehuis}, \bibinfo{person}{N~Nelly
  Litvak}, {and} \bibinfo{person}{LR Waltman}.}
  \bibinfo{year}{2012}\natexlab{}.
\newblock \showarticletitle{Predicting the long-term citation impact of recent
  publications}.
\newblock \bibinfo{journal}{\emph{Journal of Informetrics}}
  \bibinfo{volume}{9} (\bibinfo{year}{2012}), \bibinfo{pages}{642--657}.
\newblock


\bibitem[\protect\citeauthoryear{Trust}{Trust}{2016}]%
        {welcom2016}
\bibfield{author}{\bibinfo{person}{Wellcome Trust}.}
  \bibinfo{year}{2016}\natexlab{}.
\newblock \bibinfo{title}{Sharing data during Zika and other global health
  emergencies}.
\newblock
\newblock
\urldef\tempurl%
\url{https://wellcome.org/news/sharing-data-during-zika-and-other-global-health-emergencies}
\showURL{%
Retrieved Octorber 11, 2020 from \tempurl}


\bibitem[\protect\citeauthoryear{Waltman, van Eck, and van Raan}{Waltman
  et~al\mbox{.}}{2012}]%
        {Waltman2012}
\bibfield{author}{\bibinfo{person}{Ludo Waltman}, \bibinfo{person}{Nees~Jan van
  Eck}, {and} \bibinfo{person}{Anthony F.~J. van Raan}.}
  \bibinfo{year}{2012}\natexlab{}.
\newblock \showarticletitle{Universality of citation distributions revisited}.
\newblock \bibinfo{journal}{\emph{Journal of the Association for Information
  Science and Technology}}  \bibinfo{volume}{63} (\bibinfo{year}{2012}),
  \bibinfo{pages}{72--77}.
\newblock


\bibitem[\protect\citeauthoryear{Xiao}{Xiao}{2020}]%
        {Xiao2020}
\bibfield{author}{\bibinfo{person}{Eva Xiao}.} \bibinfo{year}{2020}\natexlab{}.
\newblock \showarticletitle{Scientific Journal Pulls Over a Dozen Papers by
  Chinese Researchers}.
\newblock \bibinfo{journal}{\emph{The Wall Street Journal}}
  (\bibinfo{date}{July} \bibinfo{year}{2020}).
\newblock
\urldef\tempurl%
\url{https://www.wsj.com/articles/scientific-journal-pulls-over-a-dozen-papers-by-chinese-researchers-11594743872}
\showURL{%
\tempurl}


\end{thebibliography}

\onecolumn
\appendix

\section{Preprint citation distribution}
\label{appendix:citdis}
\begin{figure}[h]
  \centering
  \includegraphics[width=0.5\linewidth]{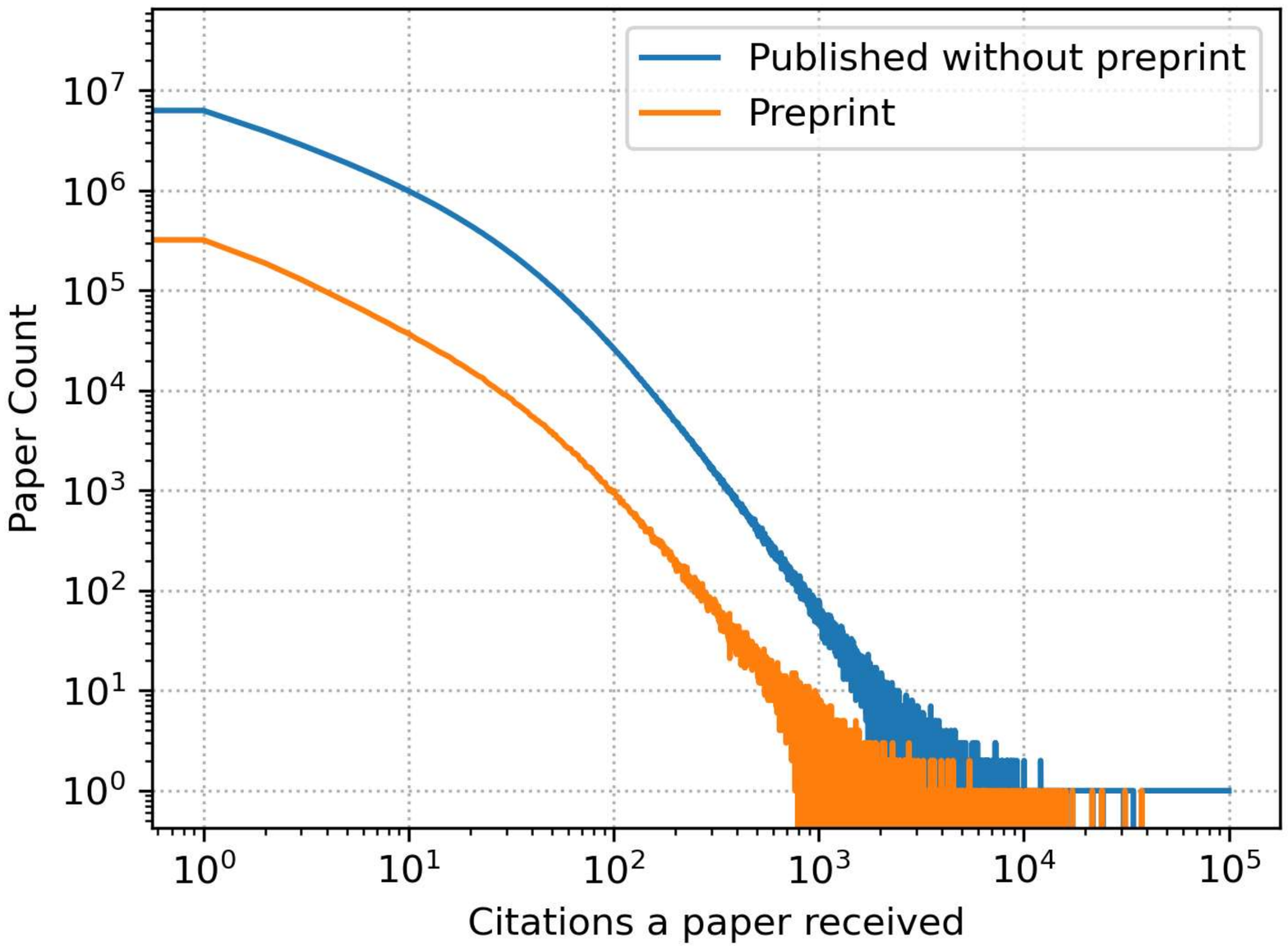}
  
\end{figure}

\section{Select journals and conferences
\label{appendix:jc}}

\begin{table*}[h]
  \label{tab:commands}
  \begin{tabular}{c|c|c|c|c|c}
    \toprule
    Journal&Domain&Paper count with preprint&Journal paper count&Preprint rate &Citation per paper\\
    \midrule
Physical Review D & Physics &61,584&87,816& 0.70& 34\\
Physical Review B & Physics &52,946&163,086& 0.32&  36 \\
The Astrophysical Journal&Physics&51,359&86,688&0.59&62\\
Physical Review Letters&Physics&40,496&104,261&0.39&82\\
Journal of Algebra & Mathematics &3,959&12,604& 0.31&  14\\
Advances in Mathematics & Mathematics &3,468&6,795& 0.51& 25\\
J. Stat. Mech. Theory Exp.& Mathematics & 3,222&4,955& 0.65&20\\
Trans Am Math Soc& Mathematics &2,992&8,094&0.37&24\\
IEEE Trans Wirel Commun & CS &1,450&9,541& 0.15&  47\\
IEEE Trans Commun& CS &1,200&10,353& 0.12&51 \\
IEEE Trans. Inf. Theory&CS&4,109 &	11,844&	0.35
&76\\
IEEE Trans. Signal Process.&CS&1,666	&15,349&	0.11
&61\\
J  Bank Financ& Economics &2,171&4,970& 0.44&88\\
Am Econ Rev& Economics &2,050&5,677& 0.36&278\\
Journal of Financial Economics&Economics&	1,969	&2,881&	0.68 &310\\
Management Science&		Economics&	1,945&	8,285&	0.23&94\\
eLife & Biology &  2,044  &  18,958  & 0.11 & 15 \\
PLoS Computational Biology & Biology &  1,339  &  8,183  & 0.16 & 43 \\
Journal of Theoretical Biology & Biology &  722  &  9,986  & 0.07 & 35 \\
PLoS Genetics&Biology&672&9,479&0.07&63\\
Scientific Reports & Multidisciplinary &5,947&119,208&0.05& 15\\
PLoS ONE & Multidisciplinary &4,551&257,798& 0.02&  24\\
Nature Communications& Multidisciplinary &4,277&33,712& 0.13&  53\\
Nature & Multidisciplinary &  1,949  &  95,790  & 0.02 & 169 \\
%Science & Multidisciplinary &  1,460  &  95,936  & 0.02 & 156 \\
  \bottomrule
\end{tabular}
    \caption{Top journals with high number of papers having preprint version (1991-2020 Sep.)}
    \label{tab:topj}
\end{table*}

\begin{table*}[h]
  \label{tab:commands}
  \begin{tabular}{c|c|c|c|c|c}
    \toprule
    Conference&Sub-domain&Paper \# with reprint& Conference paper \# & Preprint rate&Citation per paper\\
    \midrule
  CVPR & Computer Vision &  4,187  &  17,065  & 0.25 & 109 \\
NeurIPS & Machine Learning &  3,273  &  10,367  & 0.32 & 92 \\
ICML & Machine Learning &  2,793  &  8,798  & 0.32 & 73 \\
AAAI & Artificial Intelligence &  2,650  &  17,874  & 0.15 & 25 \\
ACL & Natural Language Processing &  2,008  &  9,906  & 0.20 & 49 \\
ICLR & Machine Learning &  1,887  &  2,557  & 0.74 & 77 \\
%ICASSP & Speech Recognition &  1,759  &  37,578  & 0.05 & 17 \\
ICCV & Computer Vision &  1,607  &  8,180  & 0.20 & 88 \\
%ICC &  &  1,275  &  36,126  & 0.04 & 11 \\
EMNLP & Natural Language Processing &  903  &  4,339  & 0.21 & 64 \\
KDD & Data Mining &  623  &  6,463  & 0.10 & 80 \\
WWW & World Wide Web, Information Retrieval &  622  &  6,751  & 0.09 & 69 \\
SIGIR & Information Retrieval &  350  &  5,903  & 0.06 & 56 \\
WSDM & Data Mining, Information Retrieval &  141  &  1,145  & 0.12 & 58 \\
  \bottomrule
\end{tabular}
    \caption{Top conferences with high number of papers having preprint version (1991-2020 Sep.)}
    \label{tab:topc}
\end{table*}

\end{spacing}
\end{document}